\documentclass[prd,nofootinbib, superscriptaddress,showpacs,11pt]{revtex4}
\usepackage[usenames,dvipsnames]{color}
\usepackage{ifpdf}
\ifpdf
  \usepackage{feynmf}
\else
   \usepackage{feynmp}
\fi
\usepackage{graphicx}
\usepackage{amsmath}
\usepackage{amsfonts}
\usepackage{amssymb}
\usepackage{dsfont}
\usepackage{dcolumn}
\usepackage{bm}
\usepackage{slashed}
\usepackage{pstricks}
\usepackage{slashed}
\usepackage{multirow}
\usepackage{array}
\usepackage{rotating}
\usepackage{xcolor}
\usepackage{epstopdf}
\usepackage{fancybox}

\newcommand{\GeV}      {~\mathrm{GeV}}

\newcommand{\beqn}{\begin{eqnarray}}
\newcommand{\eeqn}{\end{eqnarray}}
\newcommand{\be}{\begin{equation}}
\newcommand{\ee}{\end{equation}}

\newcommand{\mathsym}[1]{{}}

\def \n34{\tilde{\chi}^{0}_{3,4}}

\def\met100{\slashed{E}_T\geq 100 \GeV}

\newcommand{\gappeq}{\mathrel{\rlap {\raise.5ex\hbox{$>$}}
{\lower.5ex\hbox{$\sim$}}}}
\newcommand{\lappeq}{\mathrel{\rlap{\raise.5ex\hbox{$<$}}
{\lower.5ex\hbox{$\sim$}}}}

\usepackage{ulem,fancyvrb}
\def\k{\kappa}
\pacs{12.60.Fr, 12.10.Dm, 14.80.Ly}

\begin{document}
\title{\Large
The Development of Supergravity Grand Unification:\\ Circa 1982-85}
\author{R. Arnowitt\footnote{Email:arnowitt@physics.tamu.edu}}
\affiliation{Department of Physics and Astronomy,  Texas A\&M University, College Station,
Texas 77843-4242,  USA}
\author{A. H. Chamseddine\footnote{Email:chams@aub.edu.lb}}
\affiliation{American University of Beirut, Physics Department, Beirut, Lebanon}
\affiliation{Institut des Hautes Etudes Scientifique (I.H.E.S.) Bures-sur-Yvette, France}
\author{Pran~Nath\footnote{Email:nath@neu.edu}}
\affiliation{Department of Physics, Northeastern University,
 Boston, MA 02115, USA}
\date{\today}
\begin{abstract}
The development in the early eighties 
of supergravity grand unified models with gravity mediated
breaking of supersymmetry,  has led to a remarkable
progress in the study of supersymmetry at colliders, in dark matter and in a
variety of other experimental searches in the intervening years since that
time. 
The purpose of this note is to review this development and describe our  construction of 
this theory
 in the period 1982-85.
\end{abstract}
\maketitle
\clearpage

\section{ INTRODUCTION}
 Supergravity grand unification (SUGRA GUT)   is currently the main framework in theory models for the exploration 
 of new physics beyond the standard model of electroweak and strong interactions.  The framework  allows  one to extrapolate data from the electroweak scale all the way up to the scale of
 grand unification $M_G \sim 2\times 10^{16}$ GeV and vice versa to
  test new ideas
 for the unification of particles and forces.  
  Thus 
  supergravity grand unification provides the framework to test new data from the LHC
  in a broad class of models.
 Supergravity grand unification was proposed by the authors in 1982  \cite{can}
  and a significant 
  development of this field has occurred since that time led by the authors and by other researchers.
  Supergravity grand unification produces a number of theoretical insights into the high energy domain
and it has had an important impact on the progress of high energy theory since its inception  which includes providing the 
leading dark matter candidate, the neutralino.   
In the following we  summarize  the major accomplishments of the SUGRA GUT and
 indicate the key elements that were needed in the development of this model 
highlighting our contribution to the original proposal of this model in 1982 and its further development.
   \\
 
The construction of supergravity grand unification was non-trivial,  requiring several key elements.
First  it was necessary to construct couplings of gauge multiplets (representing the gauge bosons), to an arbitrary number of chiral multiples (representing the quarks and leptons) interacting supersymmetrically among themselves and to supergravity. Since spontaneous breaking of supergravity requires a super Higgs
field whose vacuum expectation value is of order of  the Planck mass ($M_{Pl}$), it was necessary to introduce
a hidden sector   (where the super Higgs fields reside) separate from  the visible sector (where the 
quarks, leptons, the electroweak Higgs and the gauge fields reside) with no direct interactions
between the hidden and the visible sectors except gravitationally.
Thus  interactions are  suppressed by factors of 
$1/M_{Pl}$ between the hidden and the visible sectors.
Such a separation of visible and hidden sector protects the low energy theory from 
mass growths of size $O(M_{Pl})$. However, there was another important gauge hierarchy problem
that needed to be resolved in supergravity grand unification due to the presence of a second 
 heavy scale, i.e., the GUT scale $M_G\simeq 2\times 10^{16}$ GeV. A priori  one could imagine corrections to the mass squared of size $m_s M_G(\k M_G)^n$ for $n = 0,1, 2,....,6$ 
 (where $m_s$ is of electroweak size and $\k =1/M_{Pl}$)  
 destroying the gauge hierarchy. Thus an important
 step in the construction of a supergravity grand unification required an explicit exhibition 
of the remarkable fact  that the
 low energy theory was free of such terms.\\

The outline of the rest of the paper is as follows:  In Sec. II we discuss global supersymmetry and the problems faced in constructing 
viable models based on global supersymmetry. In Sec. III we discuss the couplings of gauge multiplets to 
matter multiplets and to supergravity.
 In Sec. IV we discuss spontaneous breaking of supergravity and 
resolving the Planck scale hierarchy problem. In Sec. V we discuss resolving 
the GUT scale hierarchy problem.
 In Sec. VI we discuss the low energy effective theory that results when one integrates out the
 heavy fields and the breaking of $SU(2)\times U(1)$ induced by the soft breaking of supergravity
 occurs
  completing the construction of SUGRA GUTs. 
 In Sec. VII we discuss the  parameter space of soft breaking. Phenomenological implications of 
 supergravity grand unification are briefly discussed in Sec. VIII including  a brief 
 discussion of the implications of the recent discovery of a low mass possible Higgs boson 
at 125 GeV.
  Further developments 
 are discussed in Sec. IX.

\section{Problems of Model Building in Global Supersymmetry}
By about 1980, the development of global supersymmetry theory (SUSY)
(which first appeared  in two dimensions\cite{pramond} and then in four dimensions\cite{golf,volkov,bzum,Wess:1974tw})
had
reached a halting point. While supersymmetry could tame the gauge hierarchy
problem ~\cite{hierarchy,dg,s}, and achieve a reasonable grand unification of the
$SU(3)\times SU(2)\times U(1)$ coupling constants ~\cite{drw1}~\cite{dg,s},
it could do this only if
supersymmetry was broken in order to grow the masses for the new SUSY particles.
The only credible way of breaking SUSY would be  through  spontaneous breaking.
However, spontaneous breaking of SUSY appeared to be very difficult. The
SUSY Hamiltonian is positive semi-definite, and so the lowest energy state
is the supersymmetric one with $E = 0$, the broken states lying above. Further,
if SUSY was not broken at the tree level, it did not break at the loop
level~\cite{weinberg,witten}. Even if SUSY could be broken spontaneously, unsatisfactory
phenomena arose, e. g. a mass spectrum in gross disagreement with experiment, and  in
violation of what was needed for the grand unification. In addition, the
breaking of a global symmetry leads to a massless Goldstone particle, in
this case a massless fermionic goldstino, that did not appear
experimentally, as it was ruled out as a possible candidate for the neutrino.
 Thus while it was possible to exhibit the type of
  SUSY breaking
terms that maintained the hierarchies, i.e.   the so-called soft breaking terms ~\cite{girgri},
a reasonable model where such terms arose naturally and with
experimentally acceptable values did not exist. Without a theoretical
origin of the breaking of SUSY, one did not have a model that could be
compared with experiment.\\

\section{Coupling of Supergravity to Matter and Gauge Multiplets}
It was as this point in 1981 that we turned to supergravity as a possible
alternative.
This was an area in which the interests of the three authors converged. Thus  two of us
(RA and PN) had worked on local supersymmetry~\cite{an75,anz} which was a precursor of the current
supergravity theory~\cite{sup2,dz} while the other author (AHC) who also worked on supergravity \cite{CW} 
 was at that time actively engaged on the reduction of higher
dimensional supergravity theories~\cite{Chamseddine:1980cp,chams2}. 
Further, by  1980 it was clear that
nature embraced local gauge theory in both the Standard Model (SM) of
strong and electro-weak interactions and also in gravity~\cite{adm}.
 Thus using a
local supersymmetry to build models was a natural idea.
At that time,
however, there were no phenomenological supergravity models of particle
interactions. The reason for this was that the theory of how to couple an arbitrary 
matter content to supergravity did not exist. In general relativity, the coupling
of matter to gravity is very simple: one just adds the matter Lagrangian,
appropriately covariantized, to the Einstein Lagrangian to obtain minimal
coupling. In supergravity things were much more complicated. A Lagrangian
with one chiral multiplet coupled to supergravity had been constructed by
Cremmer et al. in 1979~\cite{Cremmer:1978hn} using the rules of   super-conformal
 tensor calculus~\cite{Ferrara:1978jt,sw}.
However, what was needed in order  to construct realistic models of nature was the
coupling of an arbitrary number of chiral multiplets (representing
quarks, leptons and Higgs particles) and  gauge multiplets to
supergravity, to describe the gauge interaction with the chiral multiplets
 and the gauge and chiral multiplets
 with the
supergravity fields.

This was a very complicated task, which was completed by us  by  early Spring of 1982
using the rules of supersymmetric tensor calculus. 
It was found that the most general set of couplings involved a superpotential $W(Z_A)$,
a K\"ahler potential $K(Z_A, Z_A^\dagger)$ and a gauge kinetic function
$f_{AB}(Z_A)$  where $Z_A$
are the bose components of the chiral multiplets.
The first two appeared
in the Lagrangian in the combination~\cite{can,Cremmer:1982en,Cremmer2,appliedsugra}
\beqn
G=-\kappa^2 K(Z,Z^{\dagger}) -ln(\frac{\kappa^6}{4}|W|^2),
\label{1}
\eeqn
where
$\k = 1/M_{Pl}$ , and  $M_{Pl} = 2.4\times 10^{18}$ GeV. The  Lagrangian contained the scalar potential
\beqn
 V=-\frac{1}{\kappa^4}e^{-G}(3+(G^{-1})^A_BG_{,A}G,^{B})
+\frac{1}{2\kappa^4}
Re(f^{-1})_{rs} D^r D^s
\label{3}
\eeqn
where $(G^{-1})^A_B$ is the inverse of  $G_{,A}^{~B}$ and  where 
 $D^r = -g_r \kappa^{-2} (G_i(T^r)_{ij} Z_j)$
along with a complex array of  terms involving gauge and fermionic interactions
involving $f_{AB}$ and $G$.
The potential $V$ reduces correctly to that of global SUSY in the limit of
$M_{Pl} \rightarrow \infty$, showing that one should be able to recapture the correct
physics of the Standard Model and it's SUSY generalization in the low energy limit (as
needed for any physically correct generalization). \\

However, the most
exciting result was  that due to the
supergravitational corrections, the potential $V$ was no longer positive,
and thus it was possible to spontaneously break
supersymmetry 
as was first observed  in  ~\cite{Cremmer:1978hn}
 (where a single chiral multiplet was coupled to supergravity).
In addition there was sufficient freedom to
 fine tune the cosmological constant to its
physical value.  This implied that the
(super) gravitational corrections play a crucial role in setting up
a viable supersymmetric theory. This  if realized in nature would be  a milestone in physical theory as it would be
 the first time that gravity impacted on particle physics in  a fundamental  and  direct way.\\

\section{Spontaneous Breaking  of  Supergravity and resolving  the Planck Scale Hierarchy Prolblem}

Although we had all the results on  $N = 1$ supergravity coupled  to
an arbitrary number of chiral multiplets and to vector muliplets
 in early spring of 1982, we did not immediately publish them\footnote{For previous reviews of the history see \cite{Chamseddine:2000nk}.}
  (The full details of that construction were given in
lectures at ICTP in 1983 and published in the Trieste Lecture series ~\cite{appliedsugra}).
Similarly in early spring of 1982  we had also obtained results on the breaking of a GUT group within
a supergravity grand unification context. Here we observed a new phenomena.  It was well known
that the breaking of a GUT group in global supersymmetry led to  degenerate vacua. However, in
a supergravity grand unification we noticed that this degeneracy was broken. Specifically, if
one of the vacua was chosen where the vacuum energy was arranged to vanish then the remaining
vacua  corresponded to lower vacuum energies. Again we did not immediately publish
 these results~ \cite{Nath:1982zq} and
soon after we became aware that similar results had been obtained independently by
Steven Weinberg\cite{Weinberg:1982id}. (Weinberg had in addition shown that each of the resultant
 vacua were in fact stable against bubble formation and thus the physically desirable vacuum state
with the $SU(3)\times SU(2)\times U(1)$ invariance was actually stable.)

The reason for not publishing our results both on the $N = 1$ supergravity coupled  to
an arbitrary number of chiral multiplets and to vector multiplets and on the lifting of the degeneracy
 by gravitational effects of the various vacuum states after the GUT symmetry was broken,
 was because we were focused on the  major theoretical issue of how to break
supersymmetry and  grow the necessary soft breaking terms.
 (The results on the lifting of the vacuum degeneracy were later mentioned in \cite{can} and for a later work
 see also \cite{ow}).
Again, the breaking would have to be spontaneous, e. g. by a super Higgs
effect, and a simple model that allowed for this was a linear term in the
super potential ~\cite{polonyi}: $W = m^2(Z + B)$ where Z was the super Higgs field
discussed earlier in \cite{Cremmer:1978hn}.
Minimizing the potential $V$ gave  $\kappa <Z> = O(1)$, and $B$ can be used to adjust
the cosmological constant. We later realized that the precise form of the super
Higgs potential was not crucial for the low energy theory provided it had
the form  $W = m^2f(\kappa Z)/\kappa$  where at the minimum  $\k<Z> = O(1)$ so  that
 $<W> =O(m^2/\k)$. This breaking of the supergravity gauge invariance then led to
the growth of mass for the gravitino and the SUSY particles
 of size $\k^2<W> = \k m^2$. This told us
two things:  First, if we chose $m_s\equiv$ $\k m^2 = O(TeV)$, 
i.e., $m\sim 10^{10}$ GeV, 
then the soft breaking
masses would be in the right scale to get grand unification. 
Second since we
were now breaking a local supersymmetry, the Goldstino problem was automatically
solved, as it would be  absorbed by growing mass for the gravitino (as
was initially suggested by Volkov and Akulov\cite{volkov} and worked out in detail in
\cite{Cremmer:1978hn}).

The fact that in SUSY the gauge coupling constants unify at the GUT scale
$M_G$ implies that one should construct a supergravity grand unified model
(SUGRA GUT) where the GUT group breaks to $SU(3)_C\times SU(2)_L\times U(1)_Y$ at
$M_G$  where $M_G\simeq 2\times 10^{16}$ GeV.
The main problem here was to see how to preserve the gauge
hierarchy for both $M_{Pl}$ and $M_G$ interactions.
Let us decompose the set of fields $Z_A$ so that $Z_A=(Z, Z_a)$ where $Z$ as above is  the
super Higgs field and $Z_a$  are the set of visible sector fields.
Thus a term in the
superpotential of the form  $Z Z_a Z_b$ would grow a Planck scale mass when $Z$ grows its VEV
$<Z> \sim M_{Pl}$, unless the coupling is strongly suppressed (a possibility we
will discuss later). We showed, however, that one
 may maintain the Planck scale hierarchy by
assuming the superpotential  to have  the form ~\cite{can}

\beqn
     W_{tot} = W(Z_a) + W_{SH}(Z).
     \label{4}
\eeqn
Here $W(Z_a)$ is the superpotential that governs the interactions of the chiral fields in the  visible
sector, i.e., quarks, leptons, and Higgs both light and heavy, and $W_{SH}(Z)$ is the superpotential in the hidden sector
which breaks supersymmetry spontaneously.
Now if gravity were absent, as is the case in global supersymmetry, the two sectors will
be totally separate  with no influence of one sector on another. However,  as a consequence of the fact
we are working in the framework of supergravity the effect of breaking in the hidden sector can be felt
in the visible sector, via the supergravity coupling of the two sectors (see Eq.(2)).
Thus the super Higgs lives in a  ``hidden" sector that communicates via the
effective potential V only  gravitationally. (This is now referred to as
gravity mediated breaking.)
The separation of the two sectors of Eq.(\ref{4}) guarantees, however, that the low energy sector
is protected against Planck scale mass growth by factors of $1/M_{Pl}$. 
This provided the first model where gravity plays a central role in particle physics.\\

{\section{ Resolving The GUT Scale Hierarchy Problem}}

The separation of the hidden sector from the physical sector in Eq.(\ref{4}) and the
size of $m^2$ in $m_s\equiv \k m^2$
guarantees the absence of Planck scale corrections to
the low energy masses. However, the
 nearness of the GUT scale to the Planck scale,
 leads to a new gauge
hierarchy problem, since one might have mass$^2$ corrections to the visible sector
fields of the type
\beqn
m_s M_G, ~m_s M_G (\k M_G), \cdots,  m_s M_G (\k M_G)^k,
\label{4b}
\eeqn
destroying the hierarchy for $k\leq 6$. To examine the GUT scale hierarchy, one may
divide the visible sector  fields  as $Z_a = (Z_{\alpha},  Z_i)$ where $Z_i$ are the heavy
fields and $Z_{\alpha} $ are the light fields.

To understand physically how the theory prevents the existence of corrections of the type in
Eq.(\ref{4b})  we first note that the
protection of the low energy mass from the Planck mass by having
supergravity break in the hidden sector can be rephrased differentially
(and slightly generalized) 
by requiring that the potential in the visible sector 
involves a hidden sector field dependence only in the combination $\bar Z \equiv \k Z$
\beqn
 W =  W(\bar Z, Z_{\alpha}, Z_i),
 \label{az}
\eeqn
where the only dependence of $Z$ in $W$ is in the combination $\bar Z$.
These conditions are of course satisfied if the super Higgs alone lives in a hidden
sector, but also allows coupling of the super Higgs to the physical sector
provided that coupling is suppressed by factors of $\k = 1/M_{Pl}$.
To protect the low energy masses from GUT scale masses after the breaking
of the grand unified gauge group, we impose a similar constraint isolating
the electroweak scale from the GUT scale:

\beqn
  W_{,\alpha i} = O(\k  m^2);  ~~W_{,\alpha \beta} = O(\k m^2)
  \label{alphai}
\eeqn

To show that these conditions are sufficient to protect the weak scale from
the GUT scale is non-trivial (as there are now three mass scales in the system, i.e.,  $M_{Pl}, M_G, m$)
and we give now a brief summary of the
results of \cite{nac}.
In the above case all terms of the type
~$m_s M_G (\k M_G)^k$ either cancel or are absent.
This was first shown in the specific model of \cite{can}, and  later shown to hold
for the more general class of SUGRA GUT models in ~\cite{nac}.
An alternative derivation was given in \cite{hlw}. 

The absence of $M_G$ from the soft breaking sector of the theory  is central
to setting up a  supergravity grand unified model.
 In order to obtain the low energy Lagrangian  in a GUT theory, i.e., 
 at scales
  $E <M_G$,
one must integrate out  both the  super Higgs fields as well as the GUT fields. This means in principle
that one uses field equations

\beqn
\frac{\partial V}{\partial Z}=0,  ~~ \frac{\partial V}{\partial Z_i} =0,
\label{A1}
\eeqn
 to solve for the super Higgs field
  $Z$ and for the superheavy field
  $Z_i$ in terms of the light fields $Z_\alpha$.
  To implement this procedure we  expand the VEVs
   of the super Higgs field, of the heavy fields and of the light fields
    in powers
  of $\k$ so that
  \beqn
  Z= Z^{-1} + Z^{(0)} + \cdots,\nonumber\\
  Z_{i}= Z_{i}^{(0)} + Z_{i}^{(1)} + Z_{i}^{(2)} + \cdots\nonumber\\
  Z_{\alpha}= Z_{\alpha}^{(1)} + Z_{\alpha}^{(2)}  + \dots,
  \eeqn
  where $Z^{(n)}$ is of order $\k^n$.
  Such a solution, in general,   would involve the GUT scale $M_G$ and one  needs to
  show that the low energy potential $V_{eff}$ defined by the above elimination process, i.e.,

  \beqn
  V_{eff}(Z_{\alpha})= V[ Z_i(Z_{\alpha}); ~Z_\alpha;  ~Z(Z_\alpha)],
  \label{A2}
  \eeqn
    is independent of the scale $M_G$.
    A procedure similar to this was followed in \cite{hlw}.
    The route followed in the analysis of ~\cite{can,nac} was to examine directly the
    field equations for the light modes where the extrema constraints arising from the
    super Higgs and the heavy fields, i.e., the constraints of Eq.(\ref{A1})
     have  already been imposed
     and then show that the resulting equations could be obtained by varying an effective potential
     $V_{eff}$, and that $V_{eff}$ to leading order $m_s^4$ was independent of $M_G$.

      To see the complexity of the problem one can consider the field equations for  $Z_A$
      (obtained by varying Eq.(\ref{3}))
    which take the form
\beqn
T_{AB} G_{B}=0.
\label{A3}
\eeqn
Here $G_A$ is given by
\beqn
G_A= W_{,A} + \frac{1}{2} \k^2 Z_A^{\dagger} W,
\label{A4}
\eeqn
and
$T_{AB}$   by
\beqn
 T_{AB} = W_{,AB} + \frac{1}{2} \k^2 (Z_A G_B + Z_B G_A) -\frac{1}{4} \k^4 Z_A Z_B W -\k^2 \delta_{AB} W,
 \label{A5}
\eeqn
where  $W_{,A}\equiv \partial W/\partial Z_A$.
In writing these quantities it is convenient to rescale the variables appearing as follows
\beqn
Z_{\alpha}= m_s z_\alpha, ~~ Z_i= M_G z_i, ~~ Z=M_{Pl} z,
\label{A7}
\eeqn
and further rescalings  so that
\beqn
W= m^2 M_{Pl} \overline{W}, ~
~G_{\alpha} = m_s^2 \overline{G}_{\alpha}, ~G_{i} = m_s^2 \overline{G}_{i}, ~~G_Z= m^2 \overline{G}_Z.
\label{A8}
\eeqn
where all the barred quantities are dimensionless.
The three equations of Eq.(\ref{A3}) for $A=Z, Z_i, Z_{\alpha}$ then take the form
    \beqn
\left [\frac{1}{m_s} {W}_{SH,ZZ} + (z \bar G_Z -{\overline W} - \frac{1}{4} z^2 \bar W)
 + \frac{1}{2} (\epsilon \delta_s  z_i  {\overline G}_i + \delta_s^2 z_{\alpha} {\overline G}_{\alpha})\right ] {\overline G}_Z\nonumber\\
 +\frac{1}{4} \epsilon \delta_s z z_i  {\overline G}_i {\overline W}
 + \frac{1}{2} \delta_S^2 (z {\overline G}_i^2 + z {\overline G}_{\alpha}^2 - \frac{1}{2} z z_{\alpha}
 {\overline G}_{\alpha} {\overline W}) =0,
 \label{gzz}
 \eeqn

      \beqn
 [W_{,ij} +\frac{M_G}{2} \left\{\delta_s^2 (z_i\bar G_j + z_j \bar G_i)  -\frac{1}{2} \epsilon \delta_s z_iz_j \bar W
 \right\}\nonumber\\
 + m_s \delta_{ij} \left\{\frac{1}{2} z \bar G_z - \bar W + \frac{1}{2} \delta_s^2 z_{\alpha} \bar G_{\alpha}
 \right\}] {\overline G}_j   \nonumber\\
  +  M_G \left[ \frac{1}{2} z_i {\overline G}^2_Z
   -\frac{1}{4} z_i z \overline{W}  {\overline G}_Z
   + \frac{1}{2} \delta_s^2 z_i ({\overline G}_{\alpha}^2 - \frac{1}{2} {\overline G}_{\alpha} \overline{ W})\right]=0,
   \label{A13a}
 \eeqn
and
  \beqn
 \left (\frac{1}{m_s} {W}_{,\alpha \beta} + \delta_{\alpha \beta} (\frac{1}{2} z \overline G_Z - \overline{W})
    + \frac{1}{2} \epsilon \delta_s \delta_{\alpha\beta} z_i \overline{G}_i
  + \frac{1}{2} \delta_s^2 (z_\alpha \overline{G}_\beta + z_\beta \overline{G}_\alpha)
  -\frac{1}{4} z_\alpha z_\beta \overline{W} \right ) \overline{G}_\beta
  \nonumber\\
  + \left (\overline{W}_{,\alpha i} \overline{G}_i + \frac{1}{2} z_{\alpha} \overline{G}_Z^2 -\frac{1}{4} z_\alpha z
  \overline{W}~\overline{G}_z \right)
  -\frac{1}{4} \epsilon \delta_s z_\alpha z_i  \overline{W} ~\overline{G}_i + \frac{1}{2} \delta_s^2 z_\alpha
  \overline{G}_i^2 =0.~~~~~~~~~~~
  \label{A10}
   \eeqn
Eqs.(\ref{gzz}, \ref{A13a}, \ref{A10}) contain two parameters of  smallness

 \beqn
 \epsilon = \k M_G  \sim 10^{-2}; ~~~\delta_s= \k m_s \sim 10^{-16}.
 \eeqn
(The work of \cite{hlw} does not distinguish between $M_{Pl}$ and $M_G$, and assumes
a large common mass $M$ for them.)
 Since $W_{,ij}$ is proportional to the heavy sector mass matrix, i.e., $W_{,ij} \sim M_G$, we see  that
Eq.(\ref{alphai})  allows solutions of Eqs.(\ref{gzz}, \ref{A13a}, \ref{A10})
 where all the dimensionless  barred and lower case
fields are $O(1)$ (with  corrections of size $\epsilon \delta_s$ and $\delta_s^2$). This is remarkable
since on dimensional grounds one would have expected $G_i\sim M_G^2$.
Eqs.(\ref{gzz}) and (\ref{A13a}) allow one to determine $Z$ and $Z_i$ in terms of $Z_{\alpha}$ and when
inserted into Eq.(\ref{A10}) gives the equation to determine $z_{\alpha}$ which is then independent of
$M_G$
(aside from small $\epsilon \delta_s, \delta^2_s$ corrections).
 After a detailed analysis, it is  shown that the resultant Eq.(\ref{A10}) is the field equation deducible
from varying an effective potential $V_{eff}$ depending only on the low energy fields $Z_{\alpha}$\cite{nac}.\\

The above  discussion exhibits several interesting features of the SUGRA GUT models:\\
\noindent
(i) As can be seen from Eq.(\ref{A10}) the supergravity interactions do produce additional couplings between
the GUT scale and the Planck scale fields and  the low mass section of the theory  which are not found in global
supersymmetry, but these pieces  do not add large (i.e., order  $M_G$ or $M_{Pl}$)
 corrections to the size of the low mass  fields and  thus
maintain the gauge hierarchy.

\noindent
(ii) The above equations allows one to calculate the size of the corrections to the low
energy spectrum from $M_G$ and $M_{Pl}$.  From Eq.(\ref{A10})
one sees that they
are of size $\epsilon \delta_s$ and $\delta^2_s$ and hence very small. Thus
the theory strongly protects the electroweak scale physics from high
scale physics, which justifies  using the electroweak SUSY
mass spectrum up to the GUT scale, as required by the renormalization
group analysis which ranges from the GUT scale down to the electroweak
scale.

(iii) The constraints Eq.(\ref{alphai}) put certain restrictions on the type of superpotential one may allow.
 Thus  the coupling of two light fields  and one heavy field such as $c_{\alpha\beta i} Z_{\alpha}Z_{\beta}Z_i$
 is allowed since  $W_{,\alpha i}$ in this case will be proportional to a $Z_{\beta}$ with VEV $O(m_s)$
 which satisfies the constraint of Eq.(\ref{alphai}). (The second condition of Eq.(\ref{alphai}) is satisfied
 provided the VEV of $Z_i$ vanishes.) Similarly,
the coupling of a light field to two heavy
 fields such as $c_{\alpha ij} Z_{\alpha}Z_iZ_j$ is  forbidden 
 unless the VEVs  of $Z_i$ and $Z_j$ vanish
 since $W_{,\alpha i}$ in this case will be
 proportional to $Z_j$.
  (However, such heavy-heavy-light couplings were later shown to
 destabilize the gauge hierarchy at the loop level if $Z_{\alpha}$ also had couplings to light
 particles ~\cite{Nilles:1982mp,nsw,sen}.)\\

\section{Low Energy Effective Theory and {\boldmath $SU(2)\times U(1)$} Breaking }

With these results  the effective potential of the low energy theory takes on a  remarkably simple form.
For the case $f_{\alpha\beta} = \delta_{\alpha\beta}$  and a flavor blind Kahler potential 
one has
\beqn
V_{eff} = |\frac{\partial \tilde W}{\partial Z_{\alpha}}|^2  +
m_0^2 Z_{\alpha}^{\dagger} Z_{\alpha} + (B_0 W^{(2)} + A_0 W^{(3)} + h.c)  + \frac{1}{2}[g_\sigma
\k^{-2}
(G_{\alpha}
(T^{\sigma} Z)_{\alpha})]^2,
\label{A14}
\eeqn
where $\tilde W$ is the superpotential containing only
quadratic and cubic functions of
 the light fields, i.e., $\tilde W(Z_{\alpha})
= W^{(2)}(Z_{\alpha}) + W^{(3)}(Z_{\alpha})$, and $m_0, A_0, B_0$ are soft breaking
parameters of size $m_s$.

 The proof  of the  absence of $M_G$ from the low energy theory
after spontaneous breaking of supersymmetry and of the GUT symmetry
for a specific GUT group
 was completed in early summer of 1982 and given in \cite{can} (submitted July 12).
 (The extension to the  general class of grand unified models was subsequently made in
  ~\cite{nac,hlw}.).
{\it Thus Ref.~\cite{can}, published in 1982, represents the first
construction of a viable supergravity GUT model.}\\

 It is interesting to note that Eq.(\ref{A14}) (along with gaugino masses discussed below) represent
 the first phenomenologically viable soft terms arising from spontaneous breaking of supersymmetry.
 Thus the spontaneous breaking
 of supergravity at the Planck scale gives rise naturally to the soft breaking needed phenomenologically.
The model of Ref.~\cite{can} assumed, for simplicity, an SU(5) GUT group,
a flat Kahler potential,
and  a gauge kinetic function  $f_{\alpha\beta}=\delta_{\alpha\beta}$.  The breaking of
the Standard Model gauge group to $SU(3)_C\times U(1)_{EM}$ was accomplished there at the tree level
by introduction of a singlet field $U$ with coupling $UH_1H_2$.
Soft breaking masses for squarks, sleptons and gauginos occurred
and their phenomenologies were investigated
~\cite{Weinberg:1982tp,can3,can4,can5,Dicus:1983cb,dnrx,Ellis:1982wr}.
Initially the gaugino masses were generated at the loop level~\cite{can5,apw,il1,il2,ilm,lajolla}
by the exchange of heavy fields but a more direct way of producing them 
exists by
giving the gauge kinetic energy function a non-trivial field dependence \cite{Cremmer:1982en,Cremmer2}.

The
most remarkable feature, however, was that it was the breaking of
supergravity in the hidden sector that caused the breaking of $SU(2)_L\times U(1)_Y$.
The above tree level breaking of $SU(2)_L\times U(1)_Y$ was later seen to be  unstable at the
loop level~\cite{Nilles:1982mp,nsw} and a
theoretically
more desirable model of breaking $SU(2)_L\times U(1)_Y$ could be realized using the
renormalization group equations (RGE).
Thus the running of the RGE
from the GUT scale to the electro-weak scale
drives one of the eigenvalues of the Higgs  (mass)$^2$ matrix negative
 causing the electro-weak symmetry breaking at
 this  scale ~\cite{apw}. Again it is the presence of  the soft terms arising from the  
 spontaneous breaking of supergravity that leads to the breaking of $SU(2)_L\times U(1)_Y$.
 (For early work within global supersymmetry see \cite{ir,inoue} 
 though here there was no theoretical deduction of soft breaking masses available to realize the
 possibility.).
 The origin of this remarkable effect is the largeness of the top quark Yukawa coupling (i.e., a large
 top mass). This is what drives the soft breaking of up quark Higgs (mass)$^2$
 (generated by the breaking of supergravity in the hidden sector)  downwards as we move from
 the GUT scale down to the electroweak scale.

The fact that the super Higgs mass scale, $m_s= \k m^2$, of the soft breaking parameters and the
scale of $SU(2)\times U1)$ breaking are comparable, i.e., both lie in the TeV region, is  a natural
consequence of the heavy top quark.
  In fact, SUGRA GUTs
was the only model  in the early eighties  that predicted the top quark to be
heavy i. e. $\gtrsim  100$ GeV ~\cite{Ellis:1982wr,apw,il1,il2,ilm}. Most significant is the fact that the unnatural
negative Higgs boson (mass)$^2$ term  needed in the Standard Model to generate spontaneous breaking
of the electroweak symmetry (and
grow masses for the W and Z  gauge bosons and  the quarks and leptons)
occurs in SUGRA GUTs in a natural way.\\

 During this period, a number of other researchers, who made interesting and important contributions, were engaged in examining supergravity theory
   and we conclude this section with a brief discussion of the relation between these papers and our work. The work of 
 Cremmer, Ferrara, Giradello and van Proeyen
 ~\cite{Cremmer:1982en}
 (submitted June 9, and published October 14) was concerned with the construction of the coupling of chiral and gauge multiplets to supergravity and the super Higgs effect.  As discussed above we had completed this analysis ourselves in early Spring 1982, but did not immediately publish it as we were most interested  in how to use this to construct a grand unified model. (Our work on the supergravity couplings
  was later published in \cite{appliedsugra}) These authors also generalized the supertrace formula, and showed that if $m << (M_WM_{\rm Pl})^{1/2}$ supergravity contributions were negligible, while when 
 $m > (M_WM_{\rm Pl})^{1/2}$  supergravity was important and a reasonable tree level scalar spectrum was obtained. However, they did not attempt to build any physical models based on this. While this work was submitted prior to our Physical Review Letter\cite{can} (July 12), we were unaware of it as we did not receive a preprint from the 
 authors before our paper was written.  It is also clear that we could not have written our July 12 paper without having worked out the details of the supergravity couplings independently.\\

The work of Barbieri, Ferrara and Savoy~\cite{bfs} (BFS) was submitted August 3  and so post dates our paper (July12). Further the authors were already aware of our work through the preprint sent earlier to 
Ferrara [The preprint distributed at that time contained minor errors in the form of the D term in the Lagrangian and the extrema of the superHiggs field. Neither of these effected the results in the paper, and the published version in Physical Review Letters\cite{can} contains no errors.].
 The work of BFS  makes use of the Cremmer et al. 82 paper~\cite{Cremmer:1982en} to couple low energy global SUSY models to supergravity. Thus no GUT theory is discussed. 
  Further, the electroweak breaking is not of super gravitational origin as it is anchored in the global  
  supersymmetric
  Fayet 
 model~\cite{Fayet:1974pd}.
   Thus the model is not really a simplified version of a GUT model. The BFS paper, however, was  of theoretical interest, as it introduces the technique of eliminating the superHiggs in the Lagrangian in terms of its VEV and the low energy fields, which exhibits clearly the origin of the soft breaking terms, a technique  which has been found useful in later works.\\

As discussed briefly before, the presence of a $UHH'$ term in the super potential (where $U$ is a singlet and $H$ and $H'$ are 
$5$ and $\bar 5$ representations of SU(5)) requires a large fine tuning to control loop corrections  ~\cite{Nilles:1982mp,nsw}. The work of Ferrara, Nanopoulos and Savoy
 ~\cite{Ferrara:1982ke}
(submitted December 20) was a useful addition to the literature on the subject at that time
in that it examined the possibility of evading this problem by using more complicated SU(5) representations, i. e. $75$, $50$ and $\overline{50}$. These authors found that by assuming various possible partial symmetries in the superpotential, one could indeed suppress the  loop corrections by using (technically natural) fine tunings. Thus models do exist where the loop problem can be controlled. However, the fact that the top quark turned out to be very heavy, has led to loss of interest in such approaches, and the general adoption of using the more elegant renormalization group approach described above, where the breaking of supersymmetry at the high scale by the superHiggs effect causes the breaking of $SU(2)\times U(1)$ at the electroweak scale, by naturally turning one of the eigenvalues of the Higgs mass$^2$ matrix  negative at the electroweak scale.\\

We conclude this brief discussion of other work at the time of our paper~\cite{can} with the paper of Ibanez 
\cite{Ibanez:1982ee}
(submitted August 3, revised August 24). This paper appeared after ours (July 12), and the author references our work and so was aware of it. This is the only other work at that time aside from our paper~\cite{can} 
which attempts to build a full supergravity grand unification model (though no model is fully worked out
and the low energy limit of the models considered is not examined).
Further, the models that the author considers are left, at the minimum of the effective Lagrangian, with a large negative cosmological constant, and the suggestion made to eliminate it in the Lagrangian by hand, thus breaking supersymmetry by hand. We believe that such models are basically unsatisfactory, and like the suggestions of
~\cite{bfns}
 in fact have not been pursued further.  However, the author does stress the importance of radiative corrections in going from the high energy scale down to the electroweak scale.  The subsequent (1983) work by Ibanez 
 and colleagues~\cite{il1,il2,ilm} along with the work of Alvarez-Gaume, Polchinski and Wise~\cite{apw}
 using the renormalization group approach for the mSUGRA model (of Eq(\ref{6}) below) made an important contribution to understanding how one may detect SUGRA particles at accelerators.


\section{Sources of Soft Breaking  and the parameter space of SUGRA GUTs}

While global supersymmetry can accommodate over 100 soft breaking
parameters (since there is no theory as to their origin in global supersymmetry), SUGRA GUTs
allows one to build simple models that are relatively natural  and with a significantly reduced number of
soft  terms. As mentioned already
 loop corrections can give rise to universal
gaugino masses ~\cite{can5,apw,lajolla}.
Gaugino masses can also be obtained at the tree level
by  including a field dependence in  $f_{\alpha\beta}$ to read
$\delta_{\alpha\beta} + \k c_{\alpha\beta Z} Z$
where $c_{\alpha\beta Z}$ is dimensionless $O(1)$~\cite{Cremmer:1982en,Ibanez:1982ee}.

The simple example considered above  and in \cite{can}
assumed a flat Kahler potential,
$K= \sum_A Z_A  Z^{\dagger}_A$ and implied a relation between $A_0$ and $B_0$.
However, the extension to a more general Kahler potential is straightforward.
Thus one may choose a non-flat but still flavor conserving Kahler
potential or a more general form of the Kahler potential  allowing
for couplings between the hidden and the visible sectors but suppressed by the Planck mass,
releasing the constraint between $A_0$ and $B_0$.
Thus  one might add  terms to the superpotential where the hidden sector field always occurs in the combination $\k Z$.
Since $\k <Z>\sim O(1)$ the gauge hierarchy will be preserved when a term of this type
multiplies
superpotential terms in the  visible sector~\cite{soniweldon}.
  (This is an example of violating the principle of separation of visible and hidden sectors,
which, however,  is acceptable here on phenomenological grounds  since the couplings
 are suppressed by the factor $\k = 1/M_{Pl}$.)
Thus consider a superpotential where couplings have dependence on
the hidden sector fields, so that $W= W^{(2)}+ W^{(3)}$ where $W^{(2)}= f_2(\k Z) g^{(2)}(Z_{\alpha})$
and $W^{(3)} = f_3(\k Z) g^{(3)}(Z_{\alpha})$ where $Z$ as before is  the hidden sector field and
$g^{(2)}(g^{(3)})$ are quadratic (cubic) in the visible sector fields $Z_{\alpha}$.
A direct computation then shows
that cross terms  between $W$ and $W_{SH}$
in the scalar potential  give additional contributions to dimension 2 and dimension 3 soft operators.
 With the additional term  
one finds
 $\Delta V_{eff} = \Delta B_0 W^{(2)} + \Delta A_0 W^{(3)} + h.c.$ where
 $\Delta B_0= \bar m_s  (ln f_2)'$ and
 $\Delta A_0= \bar m_s  (ln f_3)'$
 where $\bar m_s= e^{\k^2 ZZ^*} \k^2 <W_{SH}> \k <Z>$.
Note that $\Delta B_0$ and $\Delta A_0$ are uncorrelated since $f_{2}(\k Z)$ and $f_3(\k Z)$ are
arbitrary functions.
 Initially it was seen that a cubic term of the type $UH_1H_2$ leads to a bilinear term
 of the type $\mu_0 H_1H_2$ after the light singlet $U$ develops a VEV, and further  $\mu_0$  is of size $\k m^2$~\cite{can}.
 A more satisfactory derivation of
such a term was later shown to  arise by a K\"ahler transformation with a holomorphic
quadratic
term of the type $cH_1H_2 + h.c.$ in the K\"ahler potential~\cite{soniweldon,gm}.
Thus one is led to a simple model with five
universal parameters at the GUT scale: $m_0, m_{1/2}, A_0, B_0, \mu_0$.
These
parameters characterize the
 way  the super Higgs field interacts with the matter fields.
 More significantly since they arise from $\k = 1/M_{Pl}$ corrections, they give
 information on Planck scale physics.
 Models of this  type offer the possibility of understanding the origin of different mass scales.

 After minimizing the effective potential at the electro-weak scale, one is
left with four parameters and one sign:
\beqn
        m_0, ~A_0, ~m_{1/2}, ~\tan\beta=<H_2>/<H_1>, ~{\rm sign}(\mu).
        \label{6}
\eeqn
The model above has been called variously:
supergravity grand unified model\cite{an}, minimal supergravity model\cite{bbo},
  CMSSM\cite{kkrw} or mSUGRA~\cite{Baer:1996kv}
  The mSUGRA model
  \footnote{Recently,  a few people
   have
  used the notation mSUGRA to mean the special model where $B_0 = A_0 - m_0$ (thus determining $\tan\beta$ after minimizing the effective potential).  We view this as unfortunate usage as the acronym mSUGRA is clearly
  defined~\cite{Baer:1996kv}  
  and has been universally used  in the literature since  to
   mean the general model of Eq(\ref{6}). }
    is particularly simple and as a consequence of that it rapidly became
 a benchmark for the investigation of the effects of   SUSY
models on various phenomena.

As noted in the beginning of this section a  flat Kahler potential will lead to a relation between $A_0$ and $B_0$.  However, 
a general Kahler potential which is  still flavor blind  relaxes this relation and results in  Eq.({\ref{6}).
 Using the parameter space of Eq.(\ref{6}) the mass spectrum
of  the superpartners of the quarks, leptons and gauge bosons were computed and
further, the  low energy  supersymmetric Lagrangian in the mass diagonal states  was worked
out~~\cite{Weinberg:1982tp,can3,can5,Dicus:1983cb}.
Specifically the chargino and neutralino mass eigenstates were
exhibited to be linear combinations of gauginos and higgsinos which were found to lead to
a rich low energy phenomenology controlled  by the relative strength of the gaugino
and higgsino components in the mass eigenstates.
It was also noted  that the decays of the sparticles will lead to events with missing energy
in models with R parity conservation
(for a review of the early work on the phenomenological   applications of supergravity models
see ~\cite{Nath:1984tb}).\\

\section{Phenomenological Applications of SUGRA GUTS}

 Having completed the discussion of the discovery of SUGRA GUT we  now give a brief summary of some of the early subsequent developments that occurred.
Thus an early applications was the
computation of the supersymmetric loop contribution to the  anomalous magnetic
moment of the muon. It was found that the supersymmetric contribution could produce
corrections  which could be as large
or larger than the Standard Model contribution ~\cite{kks,Yuan:1984ww}. This work in part
provided the impetus for mounting the Brookhaven experiment E821. Current data
gives $a_{\mu}^{exp} - a_{\mu}^{SM} \simeq (2.87\pm 0.8) \times 10^{-9}$ 
 ($\sim 3.5\sigma$) using $e^+e^-$ annihilation for hadronic corrections 
 and using $\tau$ data one has $a_{\mu}^{exp} - a_{\mu}^{SM} \simeq (1.95\pm 0.83) \times 10^{-9}$ ($\sim 2.4\sigma$)
deviation from the Standard Model~\cite{Hoecker:2010qn,Hagiwara:2011af}.
More recent analyses indicate that supersymmetric contributions at two loop order can introduce
a correction on the order of $10\times 10^{-10}$ ~\cite{Heinemeyer:2004yq}.

Another early application of the supergravity models  to electroweak loop phenomena
was the computation of the supersymmetric electroweak corrections to the $\rho$ parameter
defined by $\rho =M_W^2/(M_Z^2 \cos^2\theta_W)$~\cite{Eliasson:1984yu}.
Corrections to this parameter are currently used to constrain supersymmetric models.
As appropriate for a rapidly developing field,  a number of further developments took place
soon after including the formulation of no-scale supergravity GUT models ~\cite{Ellis:1984bm},
and several reviews  appeared subsequent to the fast
moving developments in the period 1982-84, e.g.
~\cite{Nilles:1983ge,Haber:1984rc}.

As discussed at length  above while the soft operators
of dimensionality $d<4$ in the low energy theory
 are independent of the GUT scale, the dimension 5 or higher
operator are very sensitive to this scale. Specifically the dimension five operators
control the supersymmetric contribution to proton decay. Such dimension five
operators arise in a supersymmetric GUT theory from the elimination of the heavy higgsino triplets
which give rise to either chirality left (LL$\tilde{\rm L}\tilde {\rm L}$) or chirality right (RR$\tilde {\rm R} \tilde
{\rm R})$
operators where two of the chiral left fields or chiral right fields are for sfermions denoted by
the tilde on top.
These operators when dressed by the exchange of chargino, neutralino and gluino
exchanges  produce baryon and lepton  number violating dimension six operators
involving four fermions with chiralities of the type LLLL, LLRR, RRRR, RRLL
and lead to proton decay~\cite{wsy,sy}.

While early  analyses on the estimates
of the proton lifetime existed~\cite{drw,enr}, the supergravity Lagrangian for baryon and lepton
number violating operators for dimension five operators and their dressing with
the full set of charginos, neutralinos and gluinos were first computed in ~\cite{pdecay,pdecay2}.
This work represented the first complete analysis of proton decay lifetime
for an $SU(5)$ model exhibiting further both the low scale and as well as the
 high scale implications of a  supergravity unified model. Several new features
 emerged from this analysis including
 the importance of the L-R mixing effects and  thus  of the  LLRR and RRLL
 as well as of RRRR  operators after the dressings,  and the possibility of
 cancellations among the dressing loop contributions from the first two generations
 and the third generation. Also a systematic analysis of a number of other decays
 modes aside from the dominant $\bar \nu K^+$ mode such as   $\bar \nu \pi^+$, $\mu^+ \pi^0$,   and
 $\mu^+  K^0$ was given. These works are now standard references
 for  comparison with experiment (see, e.g., \cite{Bueno:2007um,rubbia}).

Another important early implication of the supergravity unified model was the
observation by Goldberg
~\cite{goldberg,krauss,ehnos} that models with R-parity invariance possessed
a natural candidate for the astronomically observed dark matter
if the lightest supersymmetric particle was neutral, specifically the
neutralino. Later it was shown using renormalization group analyses
that over most of the parameter space of the mSUGRA model the lightest
neutralino was indeed the lightest supersymmetric particle. Further, detailed
calculations showed the remarkable result
 that the amount of  dark matter remaining after freezeout was
 of the size that was seen astronomically. As a consequence of this,
SUGRA GUT allows one to build a model of nature ranging from the low
energy electroweak
 domain up to the GUT scale of $10^{16}$ GeV, and
  the prediction of cold dark matter within this model allows one to extrapolate backward
  in time     to
 the freeze out at
$10^{-8}$ sec.
 after the Big Bang.
 Thus the model allows for
   a remarkable unification of particle physics
with early universe cosmology.
The Large Hadron Collider can test the mSUGRA model by a study of
the lepton and jet signatures and missing energy (For a recent discussion see \cite{fln}).
Models of this type also allow
measurements of neutralino production at the LHC and
using the measurements only at the LHC
predict\cite{adgkkt}
 the amount of  dark matter
measured by Wilkinson Microwave Anisotropy Probe (WMAP) ~\cite{WMAP,WMAP2,WMAP3}
 and the Planck Satellite experiment~\cite{Bouchet:2007zz}.
 The models also predict what properties might be seen by  the
direct detection of local Milky Way  dark matter in experiments such as
the Cryogenic Dark Matter Search (CDMS)~\cite{cdms,cdms2} and the XENON100~\cite{xenon}.\\

Another early application was in the analysis of supersymmetric signals under the assumption of R parity
conservation. Thus with R parity, the decays of supersymmetric particles contain the LSP in its final
products. If the LSP was a neutral particle, such as the lightest neutralino, one will have missing  energy signatures in this case.
One of the signatures that was discussed in this context was the trileptonic one~\cite{Weinberg:1982tp,can3,can4,can5,Dicus:1983cb}
 where, for example, a chargino plus the second 
lightest neutralino ($\tilde \chi_2^0$) would have decays such as $\tilde\chi^{-}_1\to {\it l}^- \bar \nu \tilde\chi_1^0$ and 
$\tilde \chi_2^0\to {\it l}^+{\it l}^- \tilde \chi_1^0$ which would lead to a trileptonic signal and there was further work on it later on
in ~\cite{Baer:1985at,bht}. These analyses were  for on -shell decays of the W boson, i.e., $W\to \tilde \chi_1^{\pm} \tilde \chi_2^0$
with further on-shell decays of $\tilde \chi_1^{\pm}$ and  $\tilde \chi_2^0$.
However, it was pointed out in ~\cite{Nath:sw} that the reach for  discovering a chargino can be significantly extended if one considers
off -shell decays of the $W$ boson, and most modern analyses utilize this feature in the analysis of the trileptonic signal
~\cite{Baer:1994nr,Bornhauser:2011ab}. In addition to the above a variety of signatures for supersymmetry were discussed in the
1980'e and 1990's. These include analyses for the production and decay of the gluino, the squarks and for the heavy Higgs bosons
that arise in SUGRA models.  The decays of the gluino and the squarks  involve cascade decays. Thus, for example, 
gluinos which may be produced at the hadron colliders by either gluon fusion or quark fusion will undergo a decay chain such
$\tilde g \to \tilde q \bar q \to \tilde \chi_i^0 q \bar q, \tilde \chi_k^{\pm} q_1\bar q_2$ etc   while the squarks may also have cascade decays
such as $\tilde q \to \tilde g q$ where $\tilde g$ decays as indicate previously. Such decays lead to multi-jet signatures, often accompanied by
charged leptons, with missing energy. More detailed discussion of such cascade decays can be found in the literature~\cite{appliedsugra,reports,bt,dgr}.
In the early 1990's  the precision measurements of the 
grand unification of the
gauge coupling constants consistent with
supersymmetric grand unification simulated a major interest in SUGRA GUT and  as mentioned earlier there were a 
number of works~\cite{Ross:1992tz},\cite{an,bbo,kkrw,bdknt}
which discussed the sparticle spectrum and other low energy supersymmetric phenomena and further work along these
lines continues to this day. \\

With the coming on line of the LHC, the parameter space of SUGRA GUT  models and specifically of mSUGRA is being
constrained\cite{atlasreach,cmsreach}. 
More recently, LHC has found evidence of  a  possible Higgs  boson with mass around 125 GeV~\cite{:2012gk,:2012gu}.
This mass is consistent with  mSUGRA which predicts a mass range for the Higgs boson to lie below $\sim 130$ GeV
 ~\cite{Akula:2011aa,eo} (see e.g., Fig.(1) of ~\cite{Akula:2011aa}).\\

\section{Further Developments }
Of course mSUGRA is just the simplest  of the SUGRA
GUT models  and there are supergravity models with non-universalities 
such as with
non-universal gaugino masses at the GUT scale, and Higgs masses different
from squark/slepton masses at the GUT scale. These can arise from non-universal choices
for $f_{\alpha \beta}$  and $K$\cite{soniweldon}.
 Which, if any, of these models
are correct is a question that will hopefully be settled at the LHC.
However, the various SUGRA models whether with universal  or
with  non-universal soft breaking are  simply 
variations on the supergravity grand
unifications theme first proposed in ~\cite{can}.
If SUGRA GUT gets experimental support, that would impart an
impetus for a fully unified theory within the  string framework. This is  so since
string theory is seen to reduce  to SUGRA GUT models at energies below the
Planck scale~\cite{Candelas:1985en}.
 Also the concept of a hidden sector first proposed in supergravity
models could be identified with one of the $E_8$ (where supersymmetry is broken) in the $E_8\times E_8$ heterotic string theory  and   gravity mediation  is often the mechanism utilized
 for the generation of soft masses.
Thus SUGRA GUT models may be a way station on the road to  a more complete unified
 theory.\\

\noindent
 {\it Acknowledgments:}
This research work is  supported in
part by the U.S. National Science Foundation (NSF) grants
PHY-070467, PHY-0757959 and PHY-0854779.


\begin{thebibliography}{999}

\bibitem{can}
  A.~H.~Chamseddine, R.~L.~Arnowitt and P.~Nath,
  Phys.\ Rev.\ Lett.\  {\bf 49}, 970 (1982).

\bibitem{pramond}
P.~Ramond,
  Phys.\ Rev.\  D {\bf 3}, 2415 (1971).

 \bibitem{golf}
 Yu A. Golfand and E.P. Likhtman, JETP Lett. {\bf 13}, 323 (1971).

 \bibitem{volkov}
  D. Volkov and
 V.P. Akulov, JETP Lett. {\bf 16}, 438 (1972).

 \bibitem{bzum}
 J. Wess and B. Zumino, Nucl. Phys. {\bf B78}, 1 (1974).

\bibitem{Wess:1974tw}
  J.~Wess, B.~Zumino,
  Nucl.\ Phys.\  {\bf B70}, 39-50 (1974).

\bibitem{hierarchy} E. Witten, Nucl. Phys.
{\bf B188}, 513
(1981).

\bibitem{dg}
  S. Dimopoulos
and H. Georgi, Nucl. Phys. {\bf B193}, 150 (1981).

\bibitem{s}
 N. Sakai, Zeit. f. Phys.
{\bf C11}, 153 (1981).

\bibitem{drw1}
  S.~Dimopoulos, S.~Raby and F.~Wilczek,
  Phys.\ Rev.\  D {\bf 24}, 1681 (1981).

\bibitem{weinberg} S. Weinberg, Phys. Lett. {\bf B62}, 111 (1976).

\bibitem{witten}
 E. Witten, Nucl.
Phys. {\bf B202}, 253
(1982).

\bibitem{girgri} L. Giradello and M. T. Grisaru, Nucl. Phys. {\bf B194} 65 (1982).

 \bibitem{an75}
  P.~Nath and R.~L.~Arnowitt,
  Phys.\ Lett.\  B {\bf 56}, 177 (1975);

\bibitem{anz}
 R.~L.~Arnowitt, P.~Nath and B.~Zumino,
  Phys.\ Lett.\  B {\bf 56}, 81 (1975).

\bibitem{sup2} D. Freedman, S. Ferrara and P. van Nieuwenhuizen, Phys. Rev. {\bf
D13}, 3214 (1976).

\bibitem{dz}
S. Deser and B. Zumino, Phys. Lett. {\bf B62}, 335 (1976).

\bibitem{CW} A. H. Chamseddine and P. C. West, 
Nucl. Phys. \textbf{B129, }39 (1977).

\bibitem{Chamseddine:1980cp}
  A.~H.~Chamseddine,
  Nucl.\ Phys.\  {\bf B185}, 403 (1981);

  \bibitem{chams2}
      A.~H.~Chamseddine,
  Phys.\ Rev.\  {\bf D24}, 3065 (1981).

\bibitem{adm}
  R.~L.~Arnowitt, S.~Deser, C.~W.~Misner,
in ``Gravitation, an Introduction to  Current Research", ed. L. Witten, J. Wiley, New York (1962).
  [gr-qc/0405109].

\bibitem{Cremmer:1978hn}
  E.~Cremmer, B.~Julia, J.~Scherk, S.~Ferrara, L.~Girardello, P.~van Nieuwenhuizen,
  Nucl.\ Phys.\  {\bf B147}, 105 (1979).

\bibitem{Ferrara:1978jt}
  S.~Ferrara, P.~van Nieuwenhuizen,
  Phys.\ Lett.\  {\bf B76}, 404 (1978).

  \bibitem{sw}
 K.~S.~Stelle, P.~C.~West,
  Phys.\ Lett.\  {\bf B77}, 376 (1978).

\bibitem{Cremmer:1982en}
  E.~Cremmer, S.~Ferrara, L.~Girardello, A.~Van Proeyen,
  Phys.\ Lett.\  B {\bf 116}, 231 (1982).

  \bibitem{Cremmer2}
  E.~Cremmer, S.~Ferrara, L.~Girardello, A.~Van Proeyen,  
  Nucl.\ Phys.\  {\bf B212}, 413 (1983).

\bibitem{appliedsugra}
   P.~Nath, R.~L.~Arnowitt and A.~H.~Chamseddine
  ``Applied N=1 Supergravity,''
World Sci., Singapore.
and Trieste Particle Phys.1983:1 (QCD161:W626:1983).

\bibitem{Chamseddine:2000nk} 
  A.~H.~Chamseddine, R.~L.~Arnowitt and P.~Nath,
  Nucl.\ Phys.\ Proc.\ Suppl.\  {\bf 101}, 145 (2001)
  [hep-ph/0102286].

\bibitem{Nath:1982zq}
  P.~Nath, R.~L.~Arnowitt and A.~H.~Chamseddine,
  Phys.\ Lett.\  B {\bf 121}, 33 (1983).

\bibitem{Weinberg:1982id}
  S.~Weinberg,
  Phys.\ Rev.\ Lett.\  {\bf 48}, 1776 (1982).

\bibitem{ow}
  B.~A.~Ovrut and J.~Wess,
  Phys.\ Lett.\  B {\bf 119}, 105 (1982).

\bibitem{polonyi}
 J .Polonyi, University of Budapest Report No. FKI-1977-93, 1977 (unpublished).

 \bibitem{nac}
 P.~Nath, R.~L.~Arnowitt and A.~H.~Chamseddine,
  Nucl.\ Phys.\  B {\bf 227}, 121 (1983).

\bibitem{hlw}
L. Hall, J. Lykken and S. Weinberg, Phys. Rev.
{\bf D27}, 2359
(1983).

\bibitem{Nilles:1982mp}
  H.~P.~Nilles, M.~Srednicki and D.~Wyler,
  Phys.\ Lett.\  B {\bf 124}, 337 (1983);

  \bibitem{nsw}
   H.~P.~Nilles, M.~Srednicki and D.~Wyler,
  Phys.\ Lett.\  B {\bf 120}, 346 (1983);

\bibitem{sen}
 A.~Sen,
  Phys.\ Rev.\  D {\bf 30}, 2608 (1984).

\bibitem{Weinberg:1982tp}
  S.~Weinberg,
  Phys.\ Rev.\ Lett.\  {\bf 50}, 387 (1983).

\bibitem{can3}
R.~L.~Arnowitt, A.~H.~Chamseddine and P.~Nath,
  Phys.\ Rev.\ Lett.\  {\bf 50}, 232 (1983).

\bibitem{can4}
A.~H.~Chamseddine, P.~Nath and R.~L.~Arnowitt,
  Phys.\ Lett.\  B {\bf 129}, 445 (1983).

\bibitem{can5}
 P.~Nath, R.~L.~Arnowitt and A.~H.~Chamseddine,
   ``Model Independent Analysis Of Low-Energy Phenomena In Supergravity Unified
  Theories,'' NUB No: 2588, HUTP-83/A077 (unpublished 1983).

\bibitem{Dicus:1983cb}
  D.~A.~Dicus, S.~Nandi, X.~Tata,
  Phys.\ Lett.\  {\bf B129}, 451 (1983).

  \bibitem{dnrx}
  D.~A.~Dicus, S.~Nandi, W.~W.~Repko, X.~Tata,
  Phys.\ Rev.\ Lett.\  {\bf 51}, 1030 (1983).

\bibitem{Ellis:1982wr}
  J.~R.~Ellis, D.~V.~Nanopoulos and K.~Tamvakis,
  Phys.\ Lett.\  B {\bf 121}, 123 (1983).

\bibitem{apw}
  L.~Alvarez-Gaume, J.~Polchinski, M.~B.~Wise,
  Nucl.\ Phys.\  {\bf B221}, 495 (1983).

  \bibitem{il1}
 L.~E.~Ibanez, C.~Lopez,
  Phys.\ Lett.\  {\bf B126}, 54 (1983).

  \bibitem{il2}
  L.~E.~Ibanez and C.~Lopez,
  Nucl.\ Phys.\ B {\bf 233}, 511 (1984).

 \bibitem{ilm}
   L.~E.~Ibanez, C.~Lopez and C.~Munoz,
  Nucl.\ Phys.\  B {\bf 256}, 218 (1985).

 \bibitem{lajolla}
  R.~L.~Arnowitt, A.~H.~Chamseddine and P.~Nath,
 ``Supergravity and  Unification,'' HUTP-83/A014, NUB 2597, Mar 1983. 34pp.,
   Workshop on Problems in Unification and Supergravity, La Jolla, CA, Jan 1983.
Published in La Jolla  Unif. Wkshp.1983:0011 (QCD161:W61:1983)

\bibitem{ir}
 L.~E.~Ibanez, G.~G.~Ross,
  Phys.\ Lett.\  {\bf B110}, 215-220 (1982).

  \bibitem{inoue}
  K. Inoue et al., Prog. Theor. Phys. {\bf 68}, 927 (1982).

 \bibitem{bfs}
 R. Barbieri, S. Ferrara and
C. A. Savoy, Phys.
Lett. {\bf B119}, 343 (1982).

\bibitem{Fayet:1974pd}
  P.~Fayet,
  Nucl.\ Phys.\  B {\bf 90}, 104 (1975).

\bibitem{Ferrara:1982ke} 
  S.~Ferrara, D.~V.~Nanopoulos and C.~A.~Savoy,
  Phys.\ Lett.\ B {\bf 123}, 214 (1983).
%

\bibitem{Ibanez:1982ee}
  L.~E.~Ibanez,
  Phys.\ Lett.\  B {\bf 118}, 73 (1982).

\bibitem{bfns}
 R.~Barbieri, S.~Ferrara, D.~V.~Nanopoulos and K.~S.~Stelle,
  Phys.\ Lett.\  B {\bf 113}, 219 (1982).

\bibitem{soniweldon}
  S.~K.~Soni and H.~A.~Weldon,
  Phys.\ Lett.\  B {\bf 126}, 215 (1983).

   \bibitem{gm}
   G.~F.~Giudice and A.~Masiero,
  Phys.\ Lett.\  B {\bf 206}, 480 (1988).

\bibitem{an}
  R.~L.~Arnowitt and P.~Nath,
  Phys.\ Rev.\ Lett.\  {\bf 69}, 725 (1992).

\bibitem{bbo}
  V.~D.~Barger, M.~S.~Berger and P.~Ohmann,
  Phys.\ Rev.\  D {\bf 49}, 4908 (1994).

\bibitem{kkrw}
  G.~L.~Kane, C.~F.~Kolda, L.~Roszkowski and J.~D.~Wells,
  Phys.\ Rev.\  D {\bf 49}, 6173 (1994).


\bibitem{Baer:1996kv} 
  H.~Baer and M.~Brhlik,
  Phys.\ Rev.\ D {\bf 55}, 3201 (1997)
  [hep-ph/9610224].



\bibitem{Nath:1984tb}
  P.~Nath, R.~L.~Arnowitt, A.~H.~Chamseddine,
  ``N=1 Supergravity Unified Theories And Their Experimental Signatures,''
  published in "Supersymmetry and Supergravity, Nonperturbative QCD",
  Springer-Verlag, ed. P. Roy and V. Singh, 1984. p113-185.

\bibitem{kks}
 D. A. Kosower, L. M. Krauss, N. Sakai, Phys. Lett. {\bf 133B}, 305(1983).

 \bibitem{Yuan:1984ww}
   T.~C.~Yuan, R.~L.~Arnowitt, A.~H.~Chamseddine and P.~Nath,
  Z.\ Phys.\  C {\bf 26} (1984) 407.

\bibitem{Hoecker:2010qn} 
  A.~Hoecker,
  Nucl.\ Phys.\ Proc.\ Suppl.\  {\bf 218}, 189 (2011)
  [arXiv:1012.0055 [hep-ph]].

\bibitem{Hagiwara:2011af} 
  K.~Hagiwara, R.~Liao, A.~D.~Martin, D.~Nomura and T.~Teubner,
  J.\ Phys.\ G G {\bf 38}, 085003 (2011)
  [arXiv:1105.3149 [hep-ph]].

\bibitem{Heinemeyer:2004yq} 
  S.~Heinemeyer, D.~Stockinger and G.~Weiglein,
  Nucl.\ Phys.\ B {\bf 699}, 103 (2004)
  [hep-ph/0405255].

\bibitem{Eliasson:1984yu}
  E.~Eliasson, Northeastern University Preprint NUB No: 2621;
  Phys.\ Lett.\  {\bf B147}, 65 (1984).

\bibitem{Ellis:1984bm}
  J.~R.~Ellis, C.~Kounnas and D.~V.~Nanopoulos,
  Nucl.\ Phys.\  B {\bf 247}, 373 (1984).

\bibitem{Nilles:1983ge}
  H.~P.~Nilles,
  Phys.\ Rept.\  {\bf 110}, 1 (1984).

\bibitem{Haber:1984rc}
  H.~E.~Haber and G.~L.~Kane,
  Phys.\ Rept.\  {\bf 117}, 75 (1985).

\bibitem{wsy} S. Weinberg, Phys. Rev. {\bf D26}, 287 (1982).

\bibitem{sy}
 N. Sakai and T.
Yanagida, Nucl. Phys.
{\bf B197}, 533 (1982).

\bibitem{drw}
S. Dimopoulos, S. Raby and F. Wilczek, Phys. Lett. {\bf
B112}, 133 (1982).

\bibitem{enr}
J. Ellis, D. V. Nanopoulos and S. Rudaz, Nucl. Phys. {\bf B202}, 43 (1982);

 \bibitem{pdecay}
  R.~L.~Arnowitt, A.~H.~Chamseddine and P.~Nath,
  Phys.\ Lett.\  B {\bf 156}, 215 (1985).

  \bibitem{pdecay2}
    P.~Nath, A.~H.~Chamseddine and R.~L.~Arnowitt,
  Phys.\ Rev.\  D {\bf 32}, 2348 (1985).

\bibitem{Bueno:2007um}
  A.~Bueno {\it et al.},
  JHEP {\bf 0704}, 041 (2007)
  [arXiv:hep-ph/0701101].

\bibitem{rubbia}
  A.~Rubbia,
  J.\ Phys.\ Conf.\ Ser.\  {\bf 171} (2009) 012020
  [arXiv:0908.1286 [hep-ph]].

\bibitem{goldberg}
 H. Goldberg, Phys.
Rev. Lett. {\bf
50}, 1419 (1983).

\bibitem{krauss}
   L.~M.~Krauss,
  Nucl.\ Phys.\  {\bf B227}, 556 (1983).

  \bibitem{ehnos}
   J. Ellis, J. S. Hagelin, D. V. Nanopoulos, K. Olive and M.
Srednicki, Nucl.
Phys. {\bf B238}, 453 (1984).

\bibitem{fln}
   D.~Feldman, Z.~Liu and P.~Nath,
  Phys.\ Rev.\ Lett.\  {\bf 99}, 251802 (2007)
  [arXiv:0707.1873 [hep-ph]].

\bibitem{adgkkt}
  R.~L.~Arnowitt, B.~Dutta, A.~Gurrola, T.~Kamon, A.~Krislock, D.~Toback,
  Phys.\ Rev.\ Lett.\  {\bf 100}, 231802 (2008).
  [arXiv:0802.2968 [hep-ph]].

 \bibitem{WMAP}
     E.~Komatsu {\it et al.}  [WMAP Collaboration],
  Astrophys.\ J.\ Suppl.\  {\bf 192}, 18 (2011).

  \bibitem{WMAP2}
  E.~Komatsu {\it et al.}  [WMAP Collaboration],
          Astrophys.\ J.\ Suppl.\  {\bf 170}, 377 (2007).

   \bibitem{WMAP3}  
  E.~Komatsu {\it et al.}  [WMAP Collaboration],  
  Astrophys.\ J.\ Suppl.\  {\bf 148}, 175 (2003).

\bibitem{Bouchet:2007zz}
  F.~R.~Bouchet [ Planck Collaboration ],
  Mod.\ Phys.\ Lett.\  {\bf A22}, 1857-1863 (2007).

\bibitem{cdms}
  Z.~Ahmed {\it et al.} [ CDMS Collaboration ],
  Phys.\ Rev.\ Lett.\  {\bf 102}, 011301 (2009).

  \bibitem{cdms2}
  Z.~Ahmed {\it et al.} [ CDMS Collaboration ],
  Science {\bf 327}, 1619 (2010).

\bibitem{xenon}
  E.~Aprile {\it et al.}  [XENON100 Collaboration],
  arXiv:1104.2549 [astro-ph.CO].

\bibitem{Baer:1985at}
H.~Baer and X.~Tata,
Phys.\ Lett.\ B {\bf 155}, 278 (1985).

\bibitem{bht}
H.~Baer, K.~Hagiwara and X.~Tata,
Phys.\ Rev.\ Lett.\  {\bf 57}, 294 (1986).

\bibitem{Nath:sw}
P.~Nath and R.~Arnowitt,
Mod.\ Phys.\ Lett.\ A {\bf 2} (1987) 331.
%

\bibitem{Baer:1994nr} 
  H.~Baer, C.~-h.~Chen, F.~Paige and X.~Tata,
  Phys.\ Rev.\ D {\bf 50}, 4508 (1994)
  [hep-ph/9404212].

\bibitem{Bornhauser:2011ab} 
  S.~Bornhauser, M.~Drees, H.~Dreiner, O.~J.~P.~Eboli, J.~S.~Kim and O.~Kittel,
  Eur.\ Phys.\ J.\ C {\bf 72}, 1887 (2012)
  [arXiv:1110.6131 [hep-ph]].


\bibitem{reports}
  S.~P.~Martin,
  {\it Perspectives on supersymmetry II}, ed. G.~L.~Kane (World Scientific, Singapore, 1998), pp. 1-153, 
  [hep-ph/9709356].

  \bibitem{bt}
H. Baer and X. Tata, {\it Weak scale supersymmetry} (Cambridge University Press, 2006).

\bibitem{dgr}
M. Drees, R. Godbole, and P. Roy, {\it Sparticles} (World Scientific, Singapore, 2004). 


\bibitem{Ross:1992tz} 
  G.~G.~Ross and R.~G.~Roberts,
  Nucl.\ Phys.\ B {\bf 377}, 571 (1992).



\bibitem{atlasreach}
 G.~Herten, 
       ``ATLAS overview'', Talk at SUSY2012-Beijing, University of Freiburg.
       
\bibitem{cmsreach}
      G.~Tonelli,
     ``CMS overview'', Talk at SUSY2012-Beijing, CERN.

\bibitem{:2012gk} 
 G.~Aad {\it et al.}  [ATLAS Collaboration],
  Phys.\ Lett.\ B {\bf 716}, 1 (2012)
  [arXiv:1207.7214 [hep-ex]].

\bibitem{:2012gu} 
S.~Chatrchyan {\it et al.}  [CMS Collaboration],
  Phys.\ Lett.\ B {\bf 716}, 30 (2012)
  [arXiv:1207.7235 [hep-ex]].



\bibitem{Akula:2011aa} 
  S.~Akula, B.~Altunkaynak, D.~Feldman, P.~Nath and G.~Peim,
  Phys.\ Rev.\ D {\bf 85}, 075001 (2012);
  [arXiv:1112.3645 [hep-ph]].

  \bibitem{eo}
    J.~Ellis and K.~A.~Olive,
  arXiv:1202.3262 [hep-ph].

\bibitem{Candelas:1985en}
  P.~Candelas, G.~T.~Horowitz, A.~Strominger and E.~Witten,
  Nucl.\ Phys.\  B {\bf 258}, 46 (1985).

\end{thebibliography}
\end{document}